\documentclass[twocolumn,showpacs,amsmath,floatfix,prb,aps,noshowpacs]{revtex4}

\DeclareMathOperator{\sech}{sech}
\usepackage{color}
\usepackage{amsmath}
\usepackage{pifont}   
\usepackage{graphicx} 
\usepackage{dcolumn}  
\usepackage{bm}       
\usepackage{amsfonts} 
\usepackage{amssymb}  
\usepackage{multirow} 
\usepackage{romannum}
\usepackage{lipsum}
\usepackage{tikz}

\usepackage[english]{babel}
\usepackage[autostyle, english = american]{csquotes}
\MakeOuterQuote{"}

\begin{document}

\newcommand{\ba}{{\bf a}}
\newcommand{\BB}{{\bf b}}
\newcommand{\bd}{{\bf d}}
\newcommand{\br}{{\bf r}}
\newcommand{\bp}{{\bf p}}
\newcommand{\bk}{{\bf k}}
\newcommand{\bg}{{\bf g}}
\newcommand{\bt}{{\bf t}}
\newcommand{\bu}{{\bf u}}
\newcommand{\bq}{{\bf q}}
\newcommand{\bG}{{\bf G}}
\newcommand{\bP}{{\bf P}}
\newcommand{\bJ}{{\bf J}}
\newcommand{\bK}{{\bf K}}
\newcommand{\bL}{{\bf L}}
\newcommand{\bR}{{\bf R}}
\newcommand{\bS}{{\bf S}}
\newcommand{\bT}{{\bf T}}
\newcommand{\bQ}{{\bf Q}}
\newcommand{\bA}{{\bf A}}
\newcommand{\bH}{{\bf H}}
\newcommand{\bX}{{\bf X}}
\newcommand{\bsig}{\boldsymbol{\sigma}}

\author{R. Gupta$^1$}
\email{reena.gupta@bristol.ac.uk}
\author{T. Saunderson$^1$}
\author{S. Shallcross$^2$}
\author{M. Gradhand$^1$}
\author{J. Quintanilla$^3$}
\author{J. Annett$^1$}
\email{James.Annett@bristol.ac.uk}
\affiliation{1 H. H. Wills Physics Laboratory, University of Bristol, Tyndall Ave, BS8-1TL, UK}
\affiliation{2 Max-Born-Institute for non-linear optics, Max-Born Strasse 2A, 12489 Berlin, Germany}
\affiliation{3 Physics of Quantum Materials, School of Physical Sciences,University of Kent, Canterbury CT2 7NH, United Kingdom}

\title{Superconducting subphase and substantial Knight shift in $Sr_{2}RuO_{4}$}

\date{\today}

\begin{abstract}
Recent nuclear magnetic resonance experiments measuring the Knight shift in $Sr_2RuO_4$ have challenged the widely accepted picture of chiral pairing in this superconductor. Here we study the implications of helical pairing on the superconducting state while comparing our results with the available experimental data on the upper critical field and Knight shift. We solve the Bogoliubov-de-Gennes equation employing a realistic three-dimensional tight-binding model that captures the experimental Fermi surface very well. In agreement with experiments we find a Pauli limiting to the upper critical field and, at low temperatures and high fields,
a second superconducting transition. These transitions which form a superconducting subphase in the H-T phase diagram are first-order in nature and merge into a single second-order transition at a bicritical point $(T^\ast,H^\ast$), for which we find (0.8~K, 2.4~T) with experiment reporting (0.8~K, $\sim$ 1.2~T) [\textit{Phys. Rev. B} \textbf{93},
184513 (2016)]. Furthermore, we find a substantial drop in the Knight shift in agreement with recent experiments.
\end{abstract}

\maketitle

\section{Introduction}

More than two decades after the discovery of superconductivity in $Sr_2RuO_4$\cite{Maeno-1994} the nature its pairing symmetry remains unsettled.
It has been speculated\cite{Rice-1995} to be a long sought metallic analogue of superfluid helium-3 ($^3$He), and the possibility of triplet superconductivity has been explored by various groups (see Ref.~\onlinecite{Mackenzie-2003,Sigrist-2005,Mackenzie-2017,Mackenzie-2020,Kallin-2012}, and references therein). Theoretically, it was found that the free energy differences between different possible pairing symmetries were so small as to be nearly degenerate, rendering it a far from trivial problem to predict the pairing symmetry \cite{Mackenzie-2017}, a situation exacerbated by the large number of symmetry-distinct superconducting order parameters\cite{Annett-1980} compatible with the body centred tetragonal structure. Distinguishing between different order parameters therefore requires experiments to be performed under very stringent conditions. An indirect approach, where one determines specific experimental signatures of each pairing symmetry, thus provides an attractive alternative route to understanding this material\cite{physics-today}.

Early experiments pointed to $Sr_2RuO_4$ being an odd-parity chiral superconductor. Specifically, measurments of the Knight shift at both O\cite{Ishida-1998} and Ru\cite{Ishida-2001} sites showed almost no drop in value under a magnetic field applied in the $x$-$y$ plane, exactly as expected for the  
chiral $p$-wave state. Confirmation of this result was found in direct measurements of the field dependent magnetic moment by neutron scattering \cite{Duffy-2000}, although the large experimental error bars implied that a small Knight shift could not be ruled out.  
The chiral $p$-wave pairing state was further supported by phase sensitive measurements\cite{science-Nelson,science-Rice} which, under inversion, reported a phase change of $\pi$ in the superconducting order parameter. The $p$-wave chiral pairing state picture was also consistent with experiments such as muon spin rotation ($\mu SR$)\cite{Luke1998} and polar Kerr rotation\cite{PhysRevLett.97.167002} which revealed the time reversal symmetry breaking (TRSB) when $Sr_2RuO_4$ enters the superconducting phase. In contrast, the surface magnetic fields or associated edge supercurrents expected in the chiral state were never observed, despite many experimental efforts\cite{PhysRevB.76.014526}. Furthermore, recent experiments on $x$-$y$ plane uniaxial strain dependence of $T_c$ did not show the expected linear change in $T_c$ for small strains, as required theoretically for a $p_x+ i p_y$ chiral state\cite{nature-2019}, raising further doubts as to the existence of chiral $p$-wave pairing in this material\cite{PhysRevLett.122.027002,PhysRevLett.123.247001,PhysRevResearch.2.032055}.

Studies of the upper critical field\cite{JPSJ.71.2839,Maeno-2000,Maeno-2000-2,sp-heat-2014,Maeno-2013,mag-2014} revealed another serious discrepancy. At low temperatures, a first-order superconducting to normal transition in the magneto-caloric effect\cite{Maeno-2013}, the specific heat\cite{sp-heat-2014} and magnetization\cite{mag-2014} was observed under a magnetic field applied in the $x$-$y$ plane, characteristic of Pauli-limiting\cite{PhysRevB.91.144513,Ramires-2017} and inconsistent with the Knight shift measurements. For about 20 years there have been a number of attempts to resolve this puzzling behaviour with little or no success. Recently, new Knight shift experiments\cite{nature-2019}, contradicting the original experiments, observed a large drop in its value below $T_c$ for $x$-$y$ plane fields, with the previously observed temperature independent Knight shift attributed to sample heating during measurement\cite{Ishida-2020}. These new measurements decisively rule out the chiral $p$-wave pairing state and instead are consistent with the helical- or singlet-pairing in the superconducting state\cite{PhysRevB.91.144513}. Furthermore, the recent observation of half-quantized fluxoids\cite{HQV1,HQV2}, which require multiple order parameters for the pairing function with both the spin and orbital degrees of freedom active, implies the possibility of spin-triplet pairing.

Here we investigate a time reversal symmetry preserving helical pairing\cite{Roising-2019,PhysRevB.91.144513,PhysRevB.77.184515,Zhang_2014,oda19} state under an in-plane magnetic field using a realistic three-dimensional (3D) tight-binding (TB) model. We focus on results from two experimental studies \cite{PhysRevB.93.184513,Ishida-2020} to probe the internal symmetry of the Cooper pairs, and report two key findings. Firstly, as in Ref.~[\onlinecite{PhysRevB.93.184513}], we find two superconducting transitions below a temperature $T^*$, as a spin-only magnetic field is applied. These transitions are first-order in nature and merge into a single, second-order transition at a bicritical point $(T^\ast,H^\ast$), for which we find (0.8~K, 2.4~T) with experiment reporting (0.8~K, $\sim$ 1.2~T)\cite{PhysRevB.93.184513}. Secondly, our Knight shift results are in good quantitative agreement with Ref.~[\onlinecite{nature-2019, Ishida-2020}]. We find a 44\% drop in its $T = 0$~K value from the normal-state value at a field of $0.7$~Tesla. Our results therefore suggest that time reversal symmetry preserving helical pairing could be the appropriate pairing symmetry to explain many of the experimental features of $Sr_2RuO_4$. Evidently, this would then require separate explanation for other phenomena that have been interpreted as evidence of TRSB, including the increased zero-field muon spin relaxation rate in the superconducting state and the Kerr effect. A discussion of this is offered towards the end of the paper. 

The remainder of this article is structured as follows. In Sec.~\ref{TB} we describe the theoretical model employed in this work. We then (Sec.~\ref{RD}) detail our results, with the presentation divided into four subsections in which we discuss the gap-function, specific heat, spin susceptibility and Knight shift, and variation of polar angle. All the calculations are performed both at fixed temperature (varying the magnetic field) and vice-versa. Thereafter, we conclude our results with a discussion of possible future research directions in Sec.~\ref{conc}.

\section{Three dimensional tight-binding model}
\label{TB}

We employ a 3D TB Hamiltonian consisting of $d_{xy}$, $d_{xz}$, and $d_{yz}$ orbitals following the approach of Ref.~[\onlinecite{James-2003}] which was previously applied to the study of chiral pairing in the superconducting state.  The model is built upon the full 3-dimensional Fermi surface consisting of three sheets, as determined experimentally\cite{Bergemann-2000}. Superconductivity is introduced into the model
by adding a minimal set of site and orbital dependent negative $U$ pairing interactions. 
By introducing horizontal nodal lines into two of the sheets of the Fermi surface, on which the gap-function vanishes, it was shown that for the chiral superconducting state the model described the experimental specific heat very well. 

It should be noted that the experimental specific heat may be captured by either horizontal or vertical line nodes, or simply deep minima on the gap function. Recent experiments are in conflict on this matter: whereas the thermal conductivity measurements show that the gap structure of $Sr_2RuO_4$ consists of vertical line nodes \cite{vertical} with no evidence of deep minima, both spin resonance in inelastic neutron scattering measurements \cite{INS-2020} and field-angle-dependent specific heat capacity measurements \cite{Field-dependent} provided evidence of horizontal line nodes.

The key difference from Ref.~[\onlinecite{James-2003}] that we introduce here is to consider a pairing interaction that leads to helical pairing (between the same spin-types) instead of chiral pairing (between the opposite spin types). This choice of helical pairing is motivated, as explained in the introduction, by new experiments\cite{nature-2019,Ishida-2020,chronister2020evidence} in which a substantial drop in the Knight shift and magnetic susceptibility\cite{Alex-2020} has been observed under a magnetic field applied parallel to the $RuO_2$ plane. 

Our effective pairing Hamiltonian is a multi-band attractive $U$ Hubbard model with an "off-site" pairing\cite{James-2003}

\begin{align}
\label{Hamiltonian}
\hat{H}&=&\sum_{ijmm'\sigma}((\varepsilon_m-\mu)\delta_{ij}\delta_{mm'}-t_{mm'}(ij))c^{\dagger}_{im\sigma}c_{jm'\sigma}\nonumber\\
&&-\frac{1}{2}\sum_{ijmm'\sigma\sigma'}U^{\sigma\sigma'}_{mm'}(ij)\hat{n}_{im\sigma}\hat{n}_{jm'\sigma'}
\end{align}
where $m$ and $m'$ stand for the three Ruthenium $t_{2g}$ orbitals $a = d_{xy}, b = d_{xz}, c = d_{yz}$ and $i$, $j$ refer to the sites of a body centered tetragonal lattice. The hopping integrals $t_{mm}(ij)$ and on-site energies $\varepsilon_m$ have been reported in Ref.~[\onlinecite {James-2003}], which were fitted to reproduce the experimentally determined Fermi surface. The off-site pairing interaction involves two interaction constants, $U_{\parallel}$ for nearest neighbours in the plane and $U_\perp$ for nearest neighbours in adjacent planes. Also, the in-plane interaction is taken finite only for the $a-a$ pairing and the out-of-plane interaction is assumed finite for the $b-b,c-c,b-c$ types of pairings written in terms of a $3\times3$ matrix 

\begin{align}
\label{U-matrix}
\hat{U}_{m,m'}=\begin{pmatrix}
U_\parallel & 0 & 0\\
0 & U_\perp & U_\perp\\
0 & U_\perp & U_\perp
\end{pmatrix},
\end{align}
with the matrix indices ordered as $a$, $b$ and $c$ orbitals. This choice was motivated by the spatial symmetries of different orbitals: the "a" orbitals are confined to the $x$-$y$ plane and hence give rise to dominant in-plane interactions whereas the "b" and "c" orbitals having only one component lying in the plane and so contribute dominantly to the out-of-plane interaction.

We do not consider spin-orbit coupling terms in the TB model Hamiltonian, motivated by the fact that for the high field properties investigated here its role will be primarily to break the degeneracy of the 4 possible helical pairing types $A_{1u}$, $A_{2u}$, $B_{1u}$, $B_{2u}$. In preliminary calculations exploring the role of SOC, our main result of the high field subphase is found to be robust.

\begin{figure*}
\includegraphics[width=.97\linewidth]{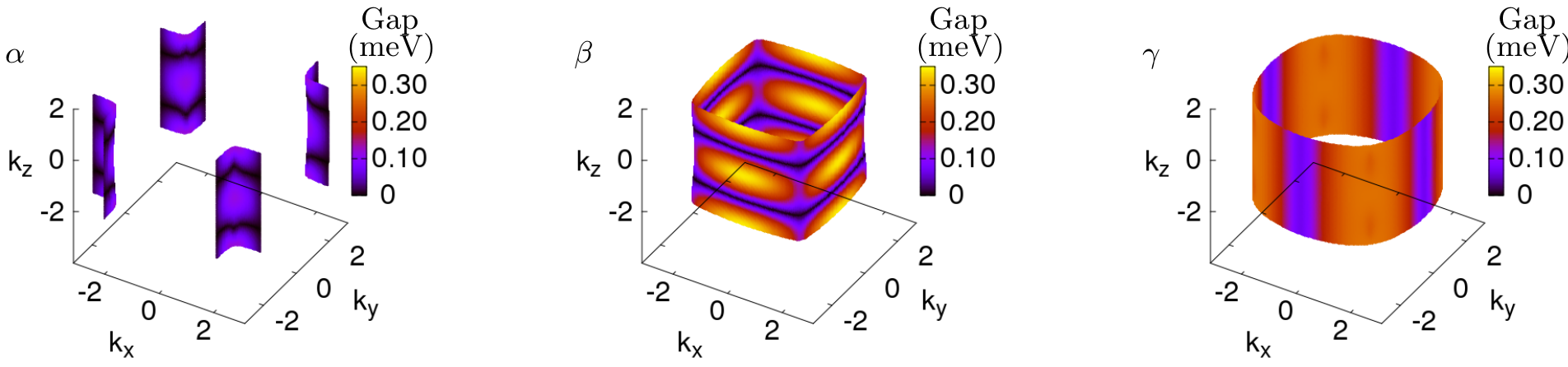}
\caption{(Colour online.) Fermi surface of $Sr_2RuO_4$ obtained from the tight-binding model described in Sec.~\ref{TB}, with the variation of superconducting gap at $T=0$~K on three Fermi sheets represented via a colour scale as indicated ($k_x, k_y$ and $k_z$ are in units of the in-plane lattice constant $a=3.862$~{\text\AA}). Horizontal line nodes are visible on the $\alpha$ and $\beta$ sheets where the gap vanishes at $k_z=\pm \pi/c$, $c=12.722$~\text\AA  ~being the lattice constant along $z$-axis.}
\label{FS-orb}
\end{figure*}

The pairing basis functions for triplet superconductivity are the odd-parity functions in $k$-space given by (where for simplicity we have chosen units of length such that the in-plane lattice constant $a=1$)

\begin{align}
\label{basis1}
\sin k_x,~~\sin k_y
\end{align}
and
\begin{align}
\label{basis2}
\sin \frac{k_x }{2}\cos \frac{k_y}{2}\cos\frac{k_z c}{2},~~\cos \frac{k_x }{2}\sin \frac{k_y}{2}\cos\frac{k_z c}{2},
\end{align}
for in-plane and out-of-plane interactions respectively.
The general form of gap-function for an odd-parity triplet state can  be represented by a $2\times2$ matrix in spin-space as

\begin{align}
\label{delta}
\hat{\Delta}(\bk)=\begin{pmatrix}
\Delta_{\uparrow\uparrow}(\bk) & \Delta_{\uparrow\downarrow} (\bk)\\
\Delta_{\downarrow\uparrow}(\bk) & \Delta_{\downarrow\downarrow}(\bk)
\end{pmatrix}
\end{align}
which can be conveniently written in the form

\begin{align}
\begin{pmatrix}
-d_x(\bk)+id_y(\bk) & d_z(\bk)\\
d_z(\bk) & d_x(\bk)+id_y(\bk)
\end{pmatrix}=i[\bd(\bk).\hat{\bsig}]\hat{\sigma}_y ,
\end{align}
where the vector ${\bd(\bk)}$ is given by ${\bd(\bk)}=(d_x(\bk),d_y(\bk),d_z(\bk))$ and $\hat{\bsig}=(\hat{\sigma}_x,\hat{\sigma}_y,\hat{\sigma}_z)$ is the vector of Pauli spin matrices.

The Bogoliubov de Gennes (BdG) equation

\begin{align}
\label{BdG}
\begin{pmatrix}
\hat{H}_{\bk}(\br) & \hat{\Delta}_{\bk}(\br)\\
\hat{\Delta}^{\dagger}_{\bk}(\br) & -\hat{H}^*_{-\bk}(\br)
\end{pmatrix}
\begin{pmatrix}
u_{n\bk}(\br)\\
v_{n\bk}(\br)
\end{pmatrix}=E_{n\bk}
\begin{pmatrix}
u_{n\bk}(\br)\\
v_{n\bk}(\br)
\end{pmatrix},
\end{align}
is solved self consistently at every $k$-point.
In our TB model, a spin-only magnetic field $\bH=(H_x,H_y,H_z)$ can be added to Eq.~\eqref{BdG} by replacing $\hat{H}_{\bk}(\br)$ with

\begin{align}
\hat{H}_{\bk}(\br)={H}_{\bk}(\br)\hat{\sigma}_0+\mu_B \mu_0 \hat{\bsig}.H,
\end{align}
$\mu_B$ being the Bohr magneton and $\mu_0$ being the vacuum permeability (in what follows we set $\mu_0=1$ for convenience).

\subsection{Pairing vector}
\label{pairing}

As $Sr_2RuO_4$ has a body-centered tetragonal crystal structure there exist several choices for the $d$-vector\cite{Annett-2006} corresponding to different irreducible representations of the point group symmetry. In this work we consider the in-plane helical $d$-vectors, of which there are four vectors corresponding to the representations $A_{1u}$, $A_{2u}$, $B_{1u}$, $B_{2u}$.  In this work we consider the form ${\bd}=(X,Y,0)$, which corresponds to the $A_{1u}$ representation. $X$ and $Y$ are the basis functions as described in Eqs.~\eqref{basis1} and ~\eqref{basis2}. We should stress that in the absence of SO coupling all four representations are degenerate and the choice of $A_{1u}$ is thus simply a representative example.

Following the approach of Ref.~[\onlinecite{Gradhand-2013}] and using Eqs.~\eqref{basis1},~\eqref{basis2} and ~\eqref{delta}, we can write expressions for the components of matrix in Eq.~\eqref{delta} as follows

\begin{align}
\Delta^{\sigma\sigma}_{aa}(\bk)=\biggl(\eta\Delta^{\sigma\sigma,x}_{aa}\sin k_x +i\Delta^{\sigma\sigma,y}_{aa}\sin k_y\biggr)
\end{align}
for in-plane components and 

\begin{figure*}
\includegraphics[width=.98\linewidth]{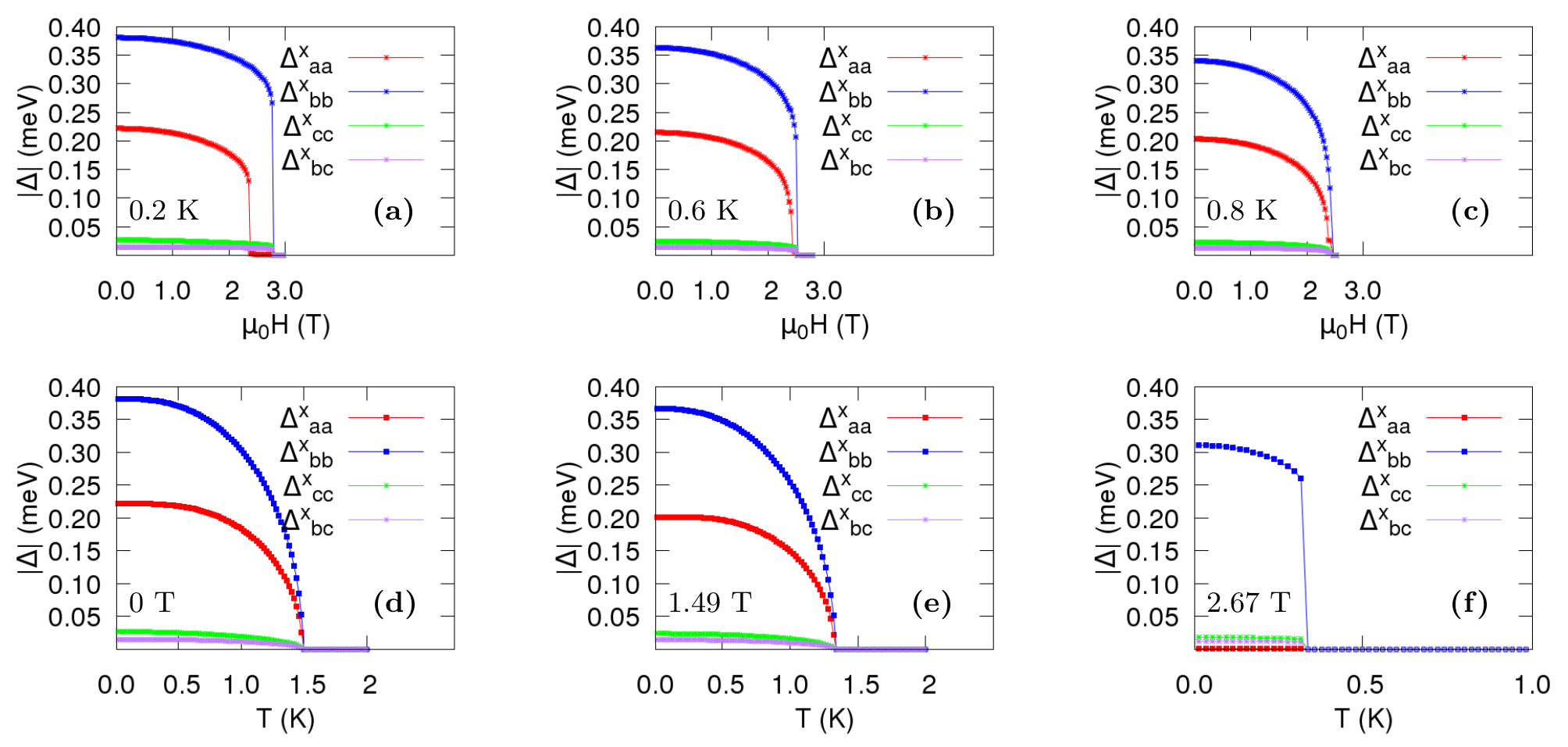}
\caption{(Colour online.) Field dependence of the gap-function at temperatures (a)~0.2 K, (b)~0.6 K, (c)~0.8 K and temperature dependence of the gap-function at fields (d)~0 T, (e)~1.49 T, (f)~2.67 T. Different plots within each panel correspond to the different components of the gap-function as labeled in the legend, where the subscripts of the gap-function denote orbitals as $a=d_{xy}$, $b=d_{xz}$, and $c=d_{yz}$. The superscript refers to the component of the gap-function; we show only the $x$ component with similar physics found for the $y$ component. Two clear first-order transitions can be seen in panels (a) and (b) at $H_{p1}$ and $H_{p2}$ that merge into a single superconducting transition in (c). The superconducting transition in (d)-(f) is of second or first order depending upon whether the field $H < H_{p1}$ or $H_{p1} < H < H_{p2}$ respectively. }
\label{gap}
\end{figure*}

\begin{align}
\Delta^{\sigma\sigma}_{ij}(\bk)=\biggl(\eta\Delta^{\sigma\sigma,x}_{ij}\sin \frac{k_x }{2}\cos \frac{k_y}{2}\nonumber\\
+i\Delta^{\sigma\sigma,y}_{ij}\cos \frac{k_x }{2}\sin \frac{k_y}{2}\biggr)\cos\frac{k_z c}{2}
\end{align}
for out-of-plane components where $ij = bb$, $cc$ and $bc$, and $\eta=+1$ for $\sigma=\downarrow$ and $\eta=-1$ for $\sigma=\uparrow$. As previously mentioned, $a=d_{xy}$, $b=d_{xz}$, and $c=d_{yz}$ represent different orbitals. The coefficients involved are given by

\begin{align}
\Delta^{\sigma\sigma,x}_{aa}=U_\parallel\times\sum_n\int d^3(\bk)[u^\sigma_{a,n}(\bk) v^{\sigma\star}_{a,n}(\bk)\nonumber\\
+v^{\sigma\star}_{a,n}(\bk) u^\sigma_{a,n}(\bk)]\times\sin k_x f(T,E_n),\nonumber\\
\Delta^{\sigma\sigma,x}_{ij}=4U_\perp\times\sum_n\int d^3(\bk)[u^\sigma_{b,n}(\bk) v^{\sigma\star}_{b,n}(\bk)\nonumber\\
+v^{\sigma\star}_{b,n}(\bk) u^\sigma_{b,n}(\bk)]\times\sin \frac{k_x }{2}\cos \frac{k_y}{2}\cos\frac{k_z c}{2}f(T,E_n),
\end{align}
where $f(T,E_n)$ is the Fermi function at a temperature $T$ and eigenvalue $E_n$ corresponding to the $n^{th}$ band.
Similar relations hold for the $y$-components $\Delta^{\sigma\sigma,y}_{aa}$ and $\Delta^{\sigma\sigma,y}_{ij}$. Using the above equations, along with the symmetry induced relations

\begin{eqnarray}
\Delta^{\sigma\sigma,x}_{aa} & = &\Delta^{\sigma\sigma,y}_{aa} \label{sym1}\nonumber\\
\Delta^{\sigma\sigma,x/y}_{bb} & = &\Delta^{\sigma\sigma,y/x}_{cc}\nonumber,
\label{sym2}
\end{eqnarray}
we self-consistently solve Eq.~\eqref{BdG}.
The only unknown constants are the in-plane ad out-of-plane interaction parameters $U_{\parallel}$ and $U_\perp$. These are chosen such that both the in-plane and out-of-plane components of the zero-field gap-function have a common superconducting critical temperature of $1.5$~K. Under this requirement we find

\begin{eqnarray}
U_\parallel&=&0.461t \\
U_\perp&=&0.624t
\end{eqnarray}
where $t=0.08162$~eV. It should be noted that in the absence of SOC the Fermi sheets are decoupled into $d_{xy}$ and $d_{yz}$/$d_{xz}$ sheets, implying that $U_\parallel$ and $U_\perp$ can be chosen independently. While this may appear artificial, implying a model of two decoupled superconductors, in Ref.~\onlinecite{James-2003} the introduction of additional subdominant interaction parameters coupling the $d_{xy}$ and $d_{yz}$/$d_{xz}$ orbitals were shown to have very little impact on either the gap function or the superconducting transition temperature. It was thus concluded that the solution of the BdG equation is not very specific to the precise details of the model parameters, but represents a generic solution valid for a range of the possible interaction parameters. Therefore while the possibility to independently tune the $d_{xy}$ and $d_{yz}$/$d_{xz}$ Fermi sheets exists, given that lifting this constraint does not significantly impact the physics of the model it does not render the mode artificial.

In Fig.~\ref{FS-orb} we illustrate the Fermi surface of $Sr_2RuO_4$ obtained from our model along with the variation of superconducting gap, obtained by solving the BdG equation self-consistently. The line nodes incorporated into the model are visible on the  $\alpha$ and $\beta$ sheets where the gap vanishes at $k_z=\pm \pi/c$, $c=12.722$~\text\AA~ being the lattice constant along $z$-axis.  These nodes are a direct consequence of the assumed interlayer pairing interaction acting among the $d_{xz}$ and $d_{yz}$ orbitals which are primarily oriented perpendicular to the plane. In contrast, the $\gamma$ sheet of the Fermi surface predominantly corresponds to the $d_{xy}$ orbital lying in the $x$-$y$ plane. The quasiparticle gap on this sheet has no nodes, but does have deep minima for ${\bf k}$ in the $(1,0,0)$ and $(0,1,0)$ directions, as shown in Fig.~\ref{FS-orb}. 

\section{Results and discussion}
\label{RD}

Using the model described in previous section, we now numerically solve the BdG equation (Eq.~\eqref{BdG}). In the following we divide our presentation of results into three subsections. In Sec.~\ref{gap1} we study the gap-function as a function of applied magnetic field for a fixed temperature, and as a function of temperature for fixed magnetic field. In this way we build up a magnetic field versus temperature phase diagram for the superconductor. In Sec.~\ref{CV1} we show the results for specific heat as a function of temperature with fixed magnetic field and vice-versa. Finally Sec.~\ref{KS1} is dedicated to the study of Knight shift and Sec.~\ref{angle1} to the variation of polar angle. In each case we carefully compare our results with experiment.

\subsection{Gap-function and phase diagram}
\label{gap1}

One of the key findings of the experiment of Ref.~[\onlinecite{PhysRevB.93.184513}] was the emergence of a superconducting subphase below $T = 0.8$~K upon variation of magnetic field. Motivated by this, we study the gap-function as a function of magnetic field (aligned along the $[100]$ direction) in Fig.~\ref{gap} panels (a)-(c), and as a function of temperature in panels (d)-(f). Different plots within each panel represent the different components of the gap-function as labeled in the legend.

\begin{figure}
  \includegraphics[width=0.98\linewidth]{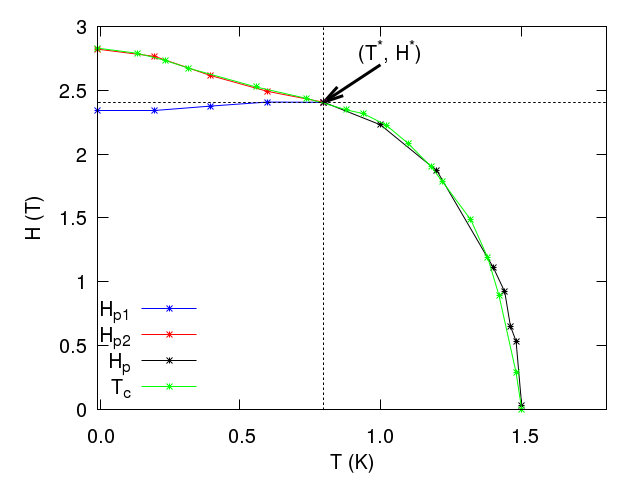} 
 \caption{(Colour online.) H-T phase diagram for $Sr_2RuO_4$ under a spin-only magnetic field $H\parallel[100]$. For $T < T^\ast$ two superconducting transitions occur with first order transitions at a lower critical field $H_{p1}$ and an upper critical field $H_{p2}$. Above this temperature a single second order superconducting transition occurs. The bicritical point $(T^*, H^*)$ at which the two phase lines merge is (0.8~K, 2.4~T), which can be compared to the experimental value (0.8~K, $\sim$1.2~T)\cite{PhysRevB.93.184513}. The line denoted by $T_c$ is the critical temperature calculated via a field sweep, and agrees to numerical precision with the $H_{p2}$ and $H_p$ lines determined from a temperature sweep at fixed field.} 
  \label{phase}
  \end{figure}

\begin{figure*}
\includegraphics[width=.98\linewidth]{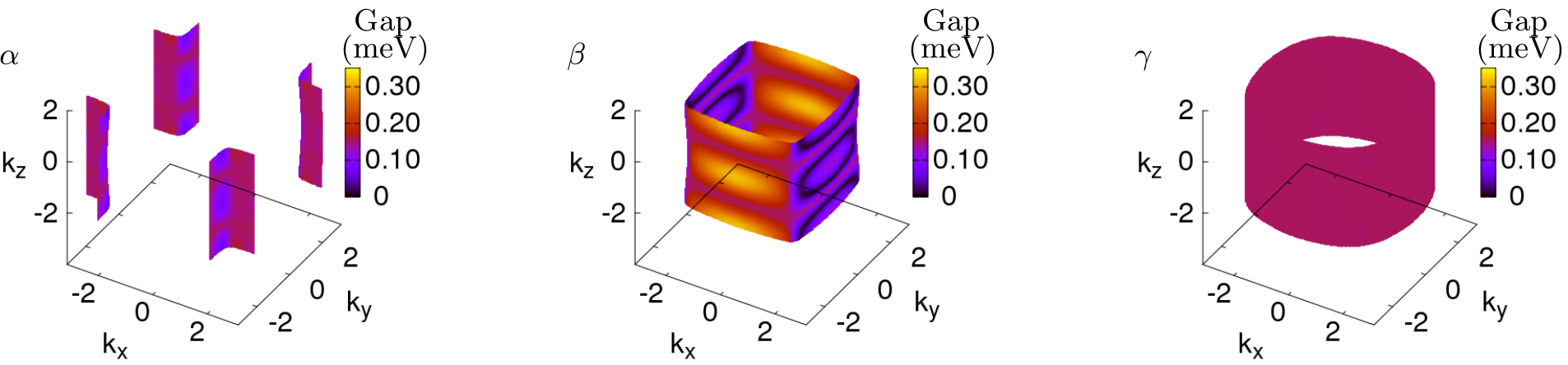}
\caption{(Colour online.) Variation of superconducting gap on the three bands comprising the Fermi surface under a magnetic field $H_x=2.67$~T ($k_x,k_y$ and $k_z$ are given in units of the in-plane lattice constant $a=3.862$~{\text\AA}). On the $\gamma$ Fermi sheet, of almost pure $d_{xy}$ character, the gap reduces to half the value found in the absence of the field (compare with Fig.~\ref{FS-orb}). Similarly, on regions of the $\alpha$- and $\beta$-sheets dominated by $d_{xz}$ orbital character (mostly along $k_y$ direction) the gap also significantly reduces. Interestingly, the nodal line structure is strikingly different from that found at zero field, as can be seen by comparison with Fig.~\ref{FS-orb}. It is to be noted that a corresponding $H_y$ field would couple to the $d_{yz}$-$d_{yz}$ pairing function and thus reduce the gap on the $d_{yz}$ dominated $k_x$ planes; note also as we employ a fully 3D model consisting of three orbitals the quantities in Fig.~\ref{gap} and Fig.~\ref{FS-gap} cannot be directly compared.}
\label{FS-gap}
\end{figure*}

\textit{Field sweep at fixed temperature}: In panel (a) we see two first order transitions at the lower critical field $H_{p1} = 2.35$~T and the upper critical field $H_{p2} = 2.77$~T, with the temperature fixed at $0.2$~K. This feature of two superconducting transitions, in our model, results from different critical fields for the gap-functions on the $d_{xy}(\Delta^x_{aa})$ and $d_{xz}/d_{yz}(\Delta^x_{bb}/\Delta^x_{cc})$ orbitals respectively, represented by $H_{p1}$ for the former and by $H_{p2}$ for the latter. It should be noted that zero temperature difference in the values of $H_{p1}$ and $H_{p2}$ in our work, which is $\sim 0.47$~T is close to the experimental value of $\sim 0.35$~T (Fig. 4(a) of Ref.\onlinecite{PhysRevB.93.184513}) for the samples with the longest average mean free path. The larger value of $H_{p2}$ implies that whereas the gap-function on $d_{xy}$ orbitals becomes zero at a lower value of the field, it remains finite on the $d_{xz}$ and $d_{yz}$ orbitals until a higher field of $H_{p2}$. When the temperature is increased to a value of 0.6~K in panel (b), the difference between $H_{p2}$ and $H_{p1}$ reduces and the two transitions move closer to each other. Upon further raising the temperature to $T = 0.8$~K, panel (c), this difference falls to zero which corresponds to a single critical field of the value $H_{p} = 2.4$~T. Above $T = 0.8$~K, the superconducting transition is of second order, which will become clearer from the specific heat results in the next section. This temperature of $0.8$~K, which we denote $T^*$, matches the temperature reported in Refs.~[\onlinecite{PhysRevB.93.184513,JPSJ.71.2839,sp-heat-2014}] below which a first-order transition has been seen. 

The first order transition is characteristic of Pauli limiting or spin limiting\cite{PhysRevB.91.144513,Ramires-2017}, also known as Chandrasekhar-Clogston limit\cite{Chandrasekhar-1962,Clogston-1962}. The paramagnetic suppression of superconductivity takes place due to the magnetic field lifting the degeneracy of electronic states with opposite momenta $\bk$ and $-\bk$ that form the Cooper pair. Pauli-limiting then occurs when the magnetic energy is larger than the condensation energy. For a singlet superconductor with an isotropic gap, the condition at $T = 0$~K is $(1/2)\chi_P H^2 = (1/2)N(0)\Delta^2$, where $\chi_P$ is the Pauli susceptibility, $H$ is the applied field, $N(0)$ is the density of states at the Fermi level and $\Delta$ is the superconducting gap. The Pauli field can be roughly approximated to be of the order of the magnetic field that correspond to the $T_c$ of the material\cite{Mackenzie-2017}, which gives a value of $2.23$~T for $T_c = 1.5$~K, close to our calculated value of $H_{p1} = 2.35$~T at $0$~K.
The paramagnetic pair-breaking is active
for spin-singlet pairing or triplet pairing with the $d$-vector locked in the basal plane\cite{PhysRevB.77.184515,oda19,book}.

\textit{Temperature sweep at fixed field}: We now consider temperature dependence of the gap function at constant field (panels (d)-(f)), where in experiment\cite{PhysRevB.93.184513} two superconducting transitions forming a superconducting subphase are again observed. However, as can be seen in panels (d)-(f) our model exhibits only a single superconducting transition temperature. Interestingly, as in experiment\cite{PhysRevB.93.184513}, we see that a continuous transition at smaller fields, panels (d) and (e), goes over to a first order transition at higher fields, panel (f).

This disagreement with experiment can be better understood by examining the phase diagram, Fig.~\ref{phase}. In this figure we show two critical fields $H_{p1}$ and $H_{p2}$, calculated from a sweep of $H$ for a fixed $T$, and the critical line $T_c$ (the green line) calculated from a sweep of $T$ for a fixed $H$. 
(The latter naturally coincides to numerical precision with $H_{p2}$ within the region of the superconducting subphase.)
The reason our model finds two superconducting transitions with variation of field but not with temperature is now clear, and results from the near zero slope of the lower critical line.
At temperatures $T<T^*$ a fixed $T$ line intersects the graph at both the fields $H_{p1}$ and $H_{p2}$ whereas, in contrast, a fixed field line intersects the graph at only one temperature, and depending upon whether $T<T^*$ or $T>T^*$ it will be a first or second order transition. The bicritical point $(T^*, H^*)$, the point on the phase diagram where the two critical fields merge into one, is (0.8~K, 2.4~T).
Seemingly, the spin-only field controls only the upper critical field as a function of temperature whereas experimental results suggest both $H_{p1}$ and $H_{p2}$ vary significantly with temperature.
  
To explore this further in Fig.~\ref{FS-gap} we display the variation of superconducting quasiparticle energy gap on three different bands of the Fermi surface under a magnetic field of $H_x=2.67$~T. Comparison with Fig.~\ref{FS-orb} reveals that the gap on the parts of the Fermi surface corresponding to the $d_{xy}$ and $d_{xz}$ orbitals is significantly reduced. On the $\gamma$ sheet, which almost purely consists of the $d_{xy}$ orbitals, it reduces to approximately half of the average value of the original gap. On parts of the $\alpha$ and $\beta$ sheets which are mainly $d_{xz}$ orbital in character, it reduces to a very small value. Interestingly the nodal structure of the field dependent quasiparticle gap shown in Fig.~\ref{FS-gap} is significantly different from the zero field case seen in Fig.~\ref{FS-orb}, especially on the $\beta$ sheet. 

It is worth pointing out at this stage that the finding of a "double superconducting transition" in the early studies of Ref.~\onlinecite{Maeno-2000,JPSJ.71.2839} was not subsequently seen in the latter studies involving much smaller (and thus possibly cleaner) samples\cite{sp-heat-2014}. However, the magnetic torque measurements of Ref.~[\onlinecite{PhysRevB.93.184513}] reported a superconducting subphase for ultra-clean samples under an applied field, very similar to the original work of Maeno et al.\cite{Maeno-2000,JPSJ.71.2839}. Interestingly, in that work the high field subphase was seen clearly only in the sample with longest mean free path, suggesting that the high-field subphase is highly sensitive to disorder. In our work, this subphase has its origin in distinct superconducting transitions on the $\gamma$ and $\alpha$, $\beta$ Fermi sheets, of $d_{xy}$ and $d_{yz}$/$d_{xz}$ orbital character respectively. We speculate that disorder that strongly couples these sheets will likely destroy this high field subphase, although we note that our preliminary SOC calculations that demonstrate subphase robustness to the orbital mixing induced by SOC suggest that very significant mixing is required to destroy the subphase. Of course, other disorder effects cannot be ruled out.

\subsection{Specific heat}
\label{CV1}
\begin{figure}[t!]
\includegraphics[width=0.98\linewidth]{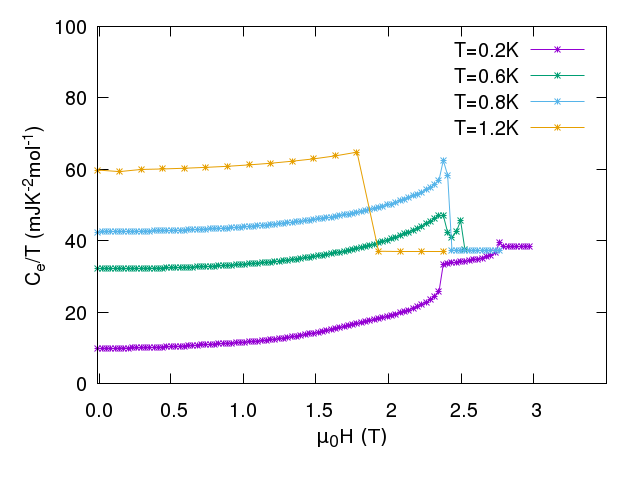}
\caption{(Colour online.) Magnetic field dependence of $C_e/T$ at various fixed temperatures. Whereas a single phase transition exists above $T^* = 0.8$~K, double superconducting transition appears below $T^*$ in well agreement with the Fig.~\ref{gap}}
\label{CV-B}
\end{figure}
Contradicting the expectation of a $C_e/T$ versus $T$ curve deviating downward near $T_c$ from the linear extrapolation of the data at lower temperatures, an unusual upward-deviation was observed at a field below $1.2$~T\cite{Maeno-2000}, while  
Ref.~[\onlinecite{Maeno-2000}] also studies $C_e/T$ versus $H$ at fixed temperature, with again a downward deviation of the $C_e/T$ versus $H$ curve near $H_{p2}$ observed at $0.5$~K and $0.7$~K and for $H\parallel$~[100], a double-peak structure was reported below $T = 0.8$~K\cite{JPSJ.71.2839}. In Fig.~\ref{CV-B} we present our results for the calculations of $C_e/T$ against $H$ for a range of temperatures. In concordance with the results for the gap-function (Fig.~\ref{gap}), we find a single phase transition above $T^* = 0.8$~K, and a double peak structure below $T^*$. As expected, our results below $T^*$ are in qualitative agreement with the experimental result\cite{Maeno-2000,JPSJ.71.2839} where we see a upward slope for $C_e/T$ versus $H$ graph near $H_{p1}$ and $H_{p2}$ at low temperatures. As mentioned in the previous section, the zero temperature difference in the values of $H_{p1}$ and $H_{p2}$ reported here of $\sim$ 0.45 ~T is close to the experimental value of $\sim$ 0.35~T (Fig. 4(a) of Ref.~[\onlinecite{PhysRevB.93.184513}]). The important difference lies in the individual values of two critical fields, with our values being larger the experimental values. This can be understood on the basis that we employ a spin-only magnetic field, and inclusion of vortex lattice will naturally reduce these field values.

One should note that the significant difference in the low field (2.3 Tesla) and high field (2.8 Tesla) jumps in heat capacity seen in Fig.~\ref{CV-B} for $T=0.2$~K, with the high field jump much smaller. This arises as the low field transition takes place on the $d_{xy}$ orbital dominated Fermi sheet that has a much more significant weight in the density of states near the Fermi energy. The absence a low field $\sqrt{H}$ behavior results from the fact that the magnetic field employed in our calculations is a spin-only magnetic field, and therefore the contribution of the vortex lattice has not been considered\cite{Volovik1993}.

\begin{figure}[!h]
\includegraphics[width=0.98\linewidth]{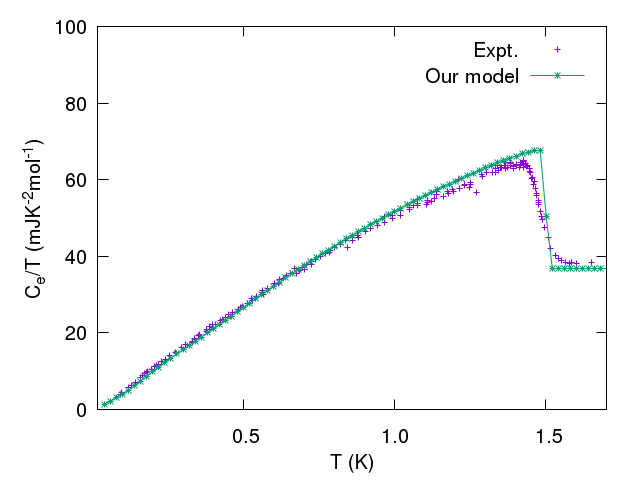}
\caption{(Colour online.) Comparison between experimentally measured~[\onlinecite{Maeno-2000}] and calculated $C_e/T$ at zero magnetic field.}
\label{CV-fig}
\end{figure}

Turning to variation of the heat capacity with temperature we first consider the zero field case, finding a very good agreement with the experimentally measured specific heat\cite{Maeno-2000} as shown Fig.~\ref{CV-fig}. The feature that at low temperature, specific heat scales linearly with $T$ is a consequence of horizontal line nodes built into our model\cite{James-2003} but, as we stress in Sec.~\ref{TB}, this linear dependence can be captured also by vertical line nodes or deep minima in the gap function.

\begin{figure}[!h]
\includegraphics[width=0.98\linewidth]{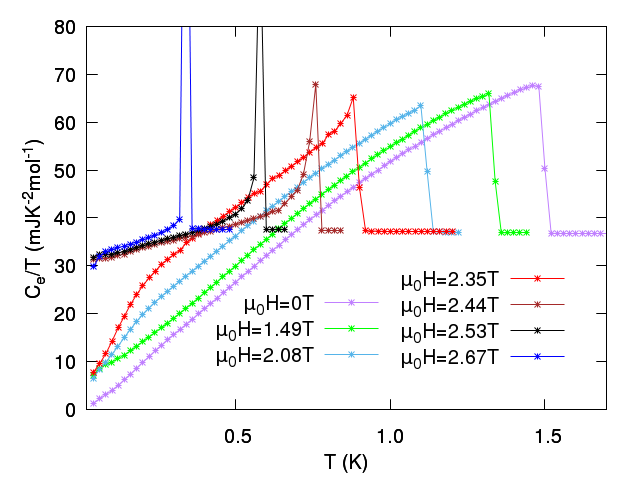}
\caption{(Colour online.) Temperature dependence of $C_e/T$ at various values of the applied field H $\parallel$ [100]. As the field is increased, a peak begins to develop at $H = 2.35$~T $\sim H^*$, characteristic of first-order transition.}
\label{CV-T}
\end{figure}

The results for the specific heat calculations at fixed magnetic field are shown in Fig.~\ref{CV-T}. 
As the field is increased, $T_c$ decreases with little change in the height of the jump until around the field $H= 2.4$~T $\sim H^*$ where the slope of the $C_e/T$ versus $T$ curve increases near $T_c$ and a peak begins to appear. This result is again in accordance with our results of the gap-function and the height of this peak increases with the increase in field. This peak is related to the Pauli paramagnetic effect\cite{JPSJ.71.2839} which results in a first-order transition and can be mathematically understood as arising from the energy derivative term, when the temperature derivative of the energy eigenvalues diverges in \cite{Leggett-1975}
\begin{align}
\label{CV}
C_v=\sum_{n,k}\biggl\{\frac{k_B \beta^2}{2}\biggl(E_{n,k}+\beta\frac{d E_k}{d\beta}\biggr)E_k{\sech}^2 \frac{\beta E_k}{2}\biggr\}.
\end{align}

\subsection{Spin susceptibility}
\label{KS1}

\begin{figure}[t!]
\centering
\begin{tikzpicture}
\draw (0, 0) node[inner sep=0] {\includegraphics[width=0.9\linewidth]{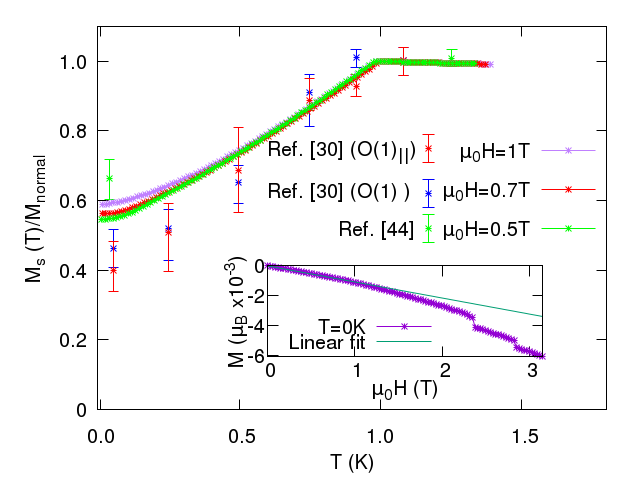}};
\draw (1, 0.5) node {$_\perp$};
\end{tikzpicture}
\caption{(Colour online.) Ratio of spin magnetic moment and the normal state moment at $0.5$~T, $0.7$~T, and $1$~T as a function of field. The ratio can be compared to the Knight shift results for spin susceptibility ratio (see text for explanation). Knight shift data from Ref.~[\onlinecite{Ishida-2020}] at $\sim 0.7$~T and polarized neutron scattering data from Ref.~[\onlinecite{Alex-2020}] $\sim 0.5$~T has also been shown for comparison. Also, shown in the inset is the field dependence of spin magnetic moment which is linear upto a field of $\sim 1.4$~T.}
\label{KS-T}
\end{figure}

The measurement of spin susceptibility has proved to be a useful technique for determination of the internal pairing state of Cooper pairs in superconductors. 
Contrary to early results \cite{Ishida-1998,Ishida-2001}, recent results report a very large drop in Knight shift\cite{nature-2019,Ishida-2020,chronister2020evidence} and in magnetic susceptibilty\cite{Alex-2020} in the superconducting state as compared to the normal state. This throws into doubt the widely accepted picture of chiral pairing in $Sr_2RuO_4$\cite{Kallin-2012} and leads to the possibility of helical pairing. As in our work we consider a magnetic field which couples only to the spin degree of freedom, we calculate a similar quantity, the spin susceptibility and compare our results with the available experimental data. We plot the ratio of spin moments in the superconducting state to the normal state in Fig.~\ref{KS-T}. We choose the values of field to be $0.7$~T from nuclear magnetic resonance (NMR)\cite{Ishida-2020} and 0.5T\cite{Alex-2020}, 1T\cite{Duffy-2000} from neutron scattering experiments performed on $Sr_2RuO_4$.

Our results can be closely compared to the NMR experiments as long as our choice of magnetic field lies in the linear-response regime so that

\begin{align}
\label{approx}
K(T)=\frac{\partial M(T)}{\partial H} = \frac{M(T)}{H}
\end{align} 
holds, where $K(T)$ is the Knight shift measured at temperature $T$ and $M(T)$ is the corresponding spin magnetic moment. As shown in the inset of Fig.~\ref{KS-T}, the linear-response holds up to a large value of the field of $\approx 1.4$~T. Our results in Fig.~\ref{KS-T} where we see a $46\%$ drop in the $T = 0$~K moment compare well with the neutron scattering results\cite{Alex-2020,Duffy-2000}. The difference with the latter could arise as neutron scattering involves the total magnetization while our calculation provides the spin only response. Also, as suggested in Ref.~[\onlinecite{Ishida-2020}], the experimental drop of a few extra percent below $50\%$ in NMR studies, a number limited by the expression for the susceptibility tensor for helical pairing\cite{Annett-2008}

\begin{align}
\label{tensor2}
\hat{\chi}_s(T)=\frac{\chi_n}{2}diag\biggl(1+Y(T),1+Y(T),2\biggr)
\end{align}
can possibly be captured by Fermi-liquid correction, where $\chi_s$, $\chi_n$ represent spin susceptibilities in the superconducting and normal state respectively and Y(T) is the Yosida function\cite{book}. 

\begin{figure}[t!]
\begin{tikzpicture}
\draw (0, 0) node[inner sep=0]
{\includegraphics[width=0.9\linewidth]{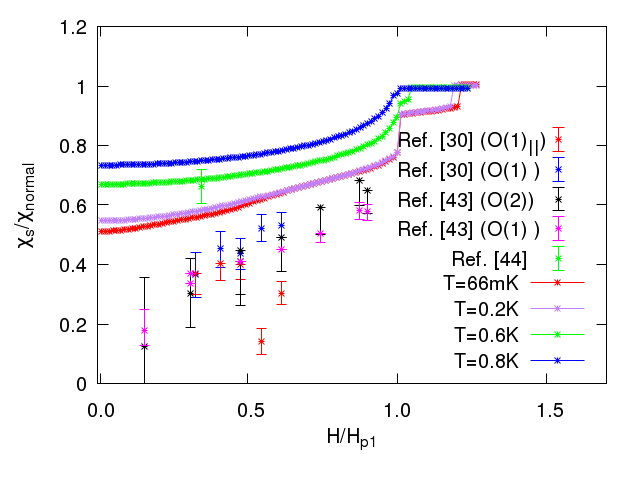}};
\draw (2.58,0.73) node {$_\perp$};
\draw (2.58,0.01) node {$_\perp$};
\end{tikzpicture}
\caption{(Colour online.) Ratio of spin susceptibilities in the superconducting and normal state at various temperatures. The result can be compared with the experimental data provided that the Eq.~\ref{approx} holds. Knight shift data from Ref.~[\onlinecite{Ishida-2020}] at $66$~mK, Ref.~[\onlinecite{chronister2020evidence}] at $25$~mK and polarized neutron scattering data from Ref.~[\onlinecite{Alex-2020}] at $0.6$~K, for purpose of comparison, has also been shown after dividing by the lower critical field values taken from Ref.~[\onlinecite{Maeno-2000}].}
\label{KS-H}
\end{figure}

Further, Refs.~[\onlinecite{Ishida-2020},\onlinecite{chronister2020evidence}] presented the Knight shift ratio in the superconducting and normal state as a function of field, at a fixed temperature of $66 $~mK. Comparing our results to these NMR measurements at oxygen site, we find some differences, especially with the [\onlinecite{chronister2020evidence}] which shows a much larger Knight shift reduction compared to the [\onlinecite{Ishida-2020}] at low field values. This could indeed imply that helical state in not the correct pairing symmetry and a spin singlet pairing is more likely. However, large error bar in the low field data of [\onlinecite{chronister2020evidence}] also does not preclude the possibility of helical pairing enhanced by Fermi liquid suppression of the susceptibility. Also, it should be noted that we cannot make a direct comparison with the oxygen NMR results within our minimal tight-binding model and it is, furthermore, likely that the O(1) site has a bigger contribution on the $\gamma$ sheet and O(2) site a bigger contribution to the $\alpha$ and $\beta$ sheets. However, a detailed analysis of these subtleties lies beyond the scope of our present manuscript.

\subsection{Varying the polar angle}
\label{angle1}

\begin{figure}[t!]
\centering
\includegraphics[width=0.98\linewidth]{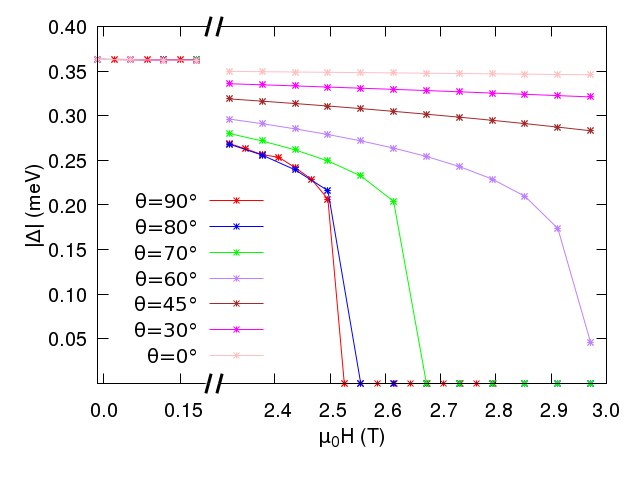}
\caption{(Colour online.) Magnetic field variation of the gap-function $|\Delta^x_{bb}|$ at $T= 0.6$~K for different field orientations with respect to the normal.}
\label{angle}
\end{figure}

Ref.~[\onlinecite{JPSJ.71.2839}] also studies the critical field by varying the polar angle between the normal to the $RuO_2$ plane and the direction of the applied magnetic field, reporting a very strong dependence on angle with $H_{p2}$ reducing sharply with the angle. This effect can not be explained by helical pairing as it is well known that a field perpendicular to the $x$-$y$ plane for a helical $d$-vector would leave the gap-function almost unchanged. In Fig.~\ref{angle}, we present the gap-function for $d_{yz}$ orbitals with a field inclined at angle $\theta$ with respect to the normal. At $\theta = 0$, when the magnetic field is out of plane, the critical field tends to infinity. As $\theta$ increases, the component of the field in the plane increases as a result of which $H_c$ decreases and becomes minimal at $\theta = 90^\circ$. A similar effect is seen for the other components of the gap-function.
Correspondingly, the Knight shift will remain unaffected for a choice of $\theta = 0^\circ$\cite{Roising-2019} (see Eq.~\ref{tensor2}).

\section{Discussion}
\label{conc}

A thorough study of helical pairing in $Sr_2RuO_4$ has been made using a realistic 3D tight-binding approach, with results compared to experiments where available. Our model based upon helical pairing agrees with many of the experimental observations such as the observation of a high field superconducting subphase, a first-order transition to the normal state, and the substantial drop of Knight shifts and magnetic moments in the superconducting phase. However, although the temperature $T^* = 0.8$~K of the bicritical point on the $H-T$ phase diagram agrees with experiment (all experimental observations of the subphase to date find $T^* = 0.8$~K), the corresponding experimental values of $H^*\sim1.2$~T and the $T=0$~K value of the lower and upper critical field 1.5~T and 1.85~T respectively do not agree, with our values for these fields being 2.4~T, 2.35~T and 2.67~T respectively. Furthermore, the temperature dependence of $H_{p1}$ also differs from experiments with our results showing a much weaker dependence.

These differences can likely be attributed to the orbital contribution to the critical field, which we do not include in our model. This will lead to some obvious differences with experiment, for example a spin-only magnetic field cannot capture the in-plane anisotropy of $H_{p1}$ and $H_{p2}$ measured via ac susceptibility studies\cite{Maeno-2002}, and so the impact on the magnitude of these fields requires further discussion. The orbital limit of the upper critical field can be estimated using the Wethamer-Helfand-Hohenberg (WHH) formula as $H^{orb}_{c2}(0) = -0.75|dH_{p2}/{dT}|_{T_c} T_c$. This formula, applied to $Sr_2RuO_4$, gives a value of $3.3$~T\cite{Lebed-2000} which would correspond to a value of $H_{p2}$ if the superconductivity was orbitally limited, significantly larger than the experimental value of $1.5$~T\cite{Maeno-2000}. This strongly indicates that the superconductivity in $Sr_2RuO_4$ is Pauli limited. Nevertheless, vortex lattice contribution to critical fields can not be ignored\cite{Agterberg-1998, Agterberg-2000, Agterberg-2005}, and may be important for obtaining quantitative agreement even in the case of Pauli limiting. Furthermore, it needs to be stressed that in our calculation we assumed that the Cooper pairs have a net zero momentum thereby excluding the possibility of FFLO phase at high field, as found, for example, in $CeCoIn_5$\cite{FFLO}, a Pauli-limited heavy-fermion superconductor. 

As discussed in the introduction to this paper, experimental evidence concerning time reversal symmetry breaking in the superconducting state of $Sr_2RuO_4$ presents a contradictory picture; our model does not support experiments which show that TRS is broken in the superconducting phase. It is, however, of interest to consider how TRSB could be recovered in the context of helical $p$-wave pairing. 
In general helical pairing states, in contrast to the the chiral pairing state, preserve TRS. This is a direct consequence of spin-orbit coupling which implies that the four states of helical type are non-degenerate:

\begin{eqnarray}
{\bd}&=&(X,Y,0)\nonumber\\
{\bd}&=&(Y,  - X,0)\nonumber\\
{\bd}&=&(X,- Y,0)\nonumber\\
{\bd}&=&(Y, X,0)
\label{a1a2b1b2}
\end{eqnarray}
each corresponding to one of the $1d$ irreducible representations $A_{1u}$, $A_{2u}$, $B_{1u}$ and $B_{2u}$ of the $D_{4h}$ point group. 
However in the absence of spin-orbit coupling they all derive from the $E_u$ irreducible representation of the tetragonal point group and among the distinct pairing states allowed are TRSB states\cite{Annett-1980}. Some of these TRSB states are the superposition of the four states in Eq.~\ref{a1a2b1b2} and in this context it is interesting to note that inclusion of SOC results in accidental or near degeneracy between pairs of the helical states above\cite{Annett-2006, PhysRevB.101.064507}. Such superposition states are worthy of future study as, in addition to possibly capturing the superconducting subphase described in this work, they may plausibly (i) yield a non-zero Kerr effect and finite orbital magnetic moment similar to those found in the chiral state\cite{Gradhand-2015, Gradhand-2017}, and (ii) may resolve the contradiction of the absence of edge super-currents, as the occurrence of such currents for TRSB helical states is unclear (in contrast to the TRSB chiral state in which they are expected).

Such pairing states are also interesting as degeneracies among helical states could explain the recently reported anomaly in the $B_{2g}$ channel\cite{ghosh2020, benhabib2020}, interpreted as indicating multiple order parameters (which of course is also consist with chiral $p$-wave, or $d+id$ pairing). Furthermore, experiments indicating possible half-flux vortices\cite{HQV1, HQV2} imply a non-abelian gauge symmetry, also requiring a multiple component order parameter. An in-plane $d$-vector as present in the helical triplet states was the first such model\cite{PhysRevB.73.220502} of half-flux vortices in $Sr_2RuO_4$. Of course other non-abelian gauge elements, such as pseudospin symmetry in orbital space\cite{Ong5486} are also possible. It is also worth mentioning that the claim that the strain experiments\cite{nature-2019} rule out multiple component order parameters is not general; while strain breaks $x$-$y$ rotational symmetry and so would split the degeneracy present in the chiral $p$-wave pairing state, it is not clear whether other degeneracies would also be lifted by strain.

In conclusion, helical pairing can explain several of the experimental features and could be a viable candidate in the search for the internal pairing symmetry of the Cooper pairs. The fact that the $A_{1u}$ pairing captures both the high field subphase as well as the suppression of knight shift suggests that variations on helical pairing (e.g. superposition states) could represent a vital further research direction. Improvements to our model include the addition of orbital contribution and SO coupling, however our preliminary calculations show that the effect of superconducting subphase is robust to the addition of the latter, as expected. 

The possibility of other types of singlet pairings such as $d$-wave or extended $s$-wave can, of course, not be ruled out\cite{Roising-2019,Sigrist-2019,Kaba-2019}, in particular since the sharp variation of $H_{p2}$ with polar angle cannot be explained with helical pairing. Furthermore, whereas the experiments [\onlinecite{nature-2019,  Ishida-2020}] showed a drop of around $50-65\%$ at $T= 0$~K, hinting towards triplet helical pairing rather than the singlet pairing -- for which a $100\%$ drop is expected -- the latest measurements on Knight shift \cite{chronister2020evidence} reports a $80-90\%$ reduction compared to the normal state at lower field values. The rather large error bar at low field value, however, does not allow one at this stage to definitively rule out a helical pairing symmetry augmented by  Fermi liquid corrections. Further experiments on the NMR measurements with a field applied along $z$-axis can help resolve the issue to some extent since no drop in Knight shift is expected for a helical pairing and such an observation would rule out any possibilities of singlet $s$- or $d$-wave pairing.

\section{Acknowledgements}
This work was carried out using the computational facilities of the Advanced Computing Research Centre, University of Bristol - http://www.bris.ac.uk/acrc/. RG, JQ, and JA acknowledge support from EPSRC through the project "Unconventional Superconductors: New paradigms for new materials" (grant references EP/P00749X/1 and EP/P007392/1). TS acknowledges support from Centre for Doctoral Training in Condensed Matter Physics, funded by EPSRC EP/L015544/1. RG expresses thanks to S. Hayden, J. Betouras, J. Buhot, and S. Ghosh for fruitful discussions.


\end{document}